\documentclass[11pt,a4paper]{article}

\usepackage{t1enc}
\usepackage[mathcal]{eucal}
\usepackage{color}
\usepackage{latexsym}
\usepackage{amsfonts}
\usepackage{amsmath}
\usepackage{amsthm}
\usepackage{amssymb}
\usepackage{mathrsfs}
\usepackage{graphicx}
\usepackage{psfrag}
\usepackage{algorithm}
\usepackage{algorithmic}

\newtheorem{theorem}{Theorem}[section]
\newtheorem{lemma}[theorem]{Lemma}

\newcommand{\prbl}{\vspace*{0.2cm}}

\newcommand{\algorithmicphase}[1]{\par\vspace*{0.3cm}\emph{#1}\vspace*{0.3cm}\par}

\begin{document}

\title{Derandomizing the Lov\'asz Local Lemma more effectively}
\author{Robin A. Moser\thanks{Research is
   supported by the SNF Grant 200021-118001/1}
\\ \vspace{0.1cm} \\
Institute for Theoretical Computer Science\\
Department of Computer Science\\
ETH Z\"urich, 8092 Z\"urich, Switzerland\\
\texttt{robin.moser@inf.ethz.ch}
}
\date{July 2008}
\maketitle

\begin{abstract} 
The famous Lov\'asz Local Lemma \cite{EL75} is a powerful tool to non-con\-struc\-ti\-ve\-ly
prove the existence of combinatorial objects meeting a prescribed collection of criteria. 
Kratochv\'il et al.~applied this technique to prove that a $k$-CNF in which
each variable appears at most $2^k/(ek)$ times is always satisfiable \cite{KST93}. In a breakthrough paper, Beck 
found that if we lower the occurrences to $\mathcal{O}(2^{k/48}/k)$, then a
deterministic polynomial-time algorithm can find a satisfying assignment to such an instance \cite{Bec91}. 
Alon randomized the algorithm and required $\mathcal{O}(2^{k/8}/k)$ occurrences \cite{Alo91}. In \cite{Mos06}, we exhibited
a refinement of his method which copes with $\mathcal{O}(2^{k/6}/k)$ of them. The hitherto
best known randomized algorithm is due to Srinivasan and is capable of solving $\mathcal{O}(2^{k/4}/k)$ 
occurrence instances \cite{Sri08}. Answering two questions asked by Srinivasan, we shall now present 
an approach that tolerates $\mathcal{O}(2^{k/2}/k)$ occurrences per variable and which can most easily be
derandomized. The new algorithm bases on an alternative type of witness tree structure and drops a number
of limiting aspects common to all previous methods.
\end{abstract}


\section{Introduction}

We assume an infinite supply of propositional \emph{variables}. A \emph{literal} $L$ is a variable
$x$ or a complemented variable $\bar x$. A finite set $D$ of literals over pairwise distinct variables
is called a \emph{clause}. We say that a variable $x$ \emph{occurs} in $D$ if $x \in D$ or $\bar x \in D$.
A finite set $\varphi$ of clauses is called a \emph{formula} or a \emph{CNF} (Conjunctive Normal Form).
We say that $\varphi$ is a $k$-\emph{CNF}, if every clause has size exactly $k$. We say that the variable 
$x$ \emph{occurs $j$ times} in a formula if there are exactly $j$ clauses in which $x$ occurs. We write 
$\mbox{var}(\varphi)$ to denote the set of all variables occurring in $\varphi$.

A \emph{truth assignment} is a function $\alpha : \mbox{var}(\varphi) \rightarrow \{0,1\}$ which assigns
a boolean value to each variable. A literal $L=x$ (or $L=\bar x$) is \emph{satisfied} by $\alpha$ 
if $\alpha(x)=1$ (or $\alpha(x)=0$). A clause is \emph{satisfied} by $\alpha$ if it contains a satisfied
literal and a formula is \emph{satisfied} if all of its clauses are. A formula is \emph{satisfiable}
if there exists a satisfying truth assignment to its variables.

In \cite{KST93}, Kratochv\'il et al.~have applied the Lov\'asz Local Lemma (from \cite{EL75}) to prove
that every $k$-CNF in which every variable occurs no more than $2^k/(ek)$ times has a satisfying assignment.
The Local Lemma currently appears to be the only workable tool for the obtention of such a bound.
Unfortunately, all known proofs based on the Local Lemma are non-constructive and do not directly allow
for the construction of an efficient algorithm to actually find a satisfying assignment. 

Such an algorithm was first provided by Beck in \cite{Bec91} and then randomized by Alon \cite{Alo91}.
Their algorithm bases on the principle of selecting a preliminary assignment and then discriminating clauses according
to whether few or many of their literals are satisfied by it. The \emph{bad} clauses containing few satisfied literals
are then reassessed and their variables are reassigned new values. Such a procedure is efficient if the dependencies
are low enough so as to guarantee that clustered components of bad clauses are very small and can be
solved in a brute force fashion. The original approach by Beck required that no variable appear a number
higher than $\mathcal{O}(2^{k/48}/k)$ of times. In Alon's simplification, the requirement was still
$\mathcal{O}(2^{k/8}/k)$.

Several authors have improved and extended those approaches, with goals somewhat complementary to what
we are concerned with. In \cite{MR98} e.g., Molloy and Reed extrapolate guidelines which allow for the construction of
an algorithmic version to numerous applications of the Local Lemma other than bounded occurrence SAT
or hypergraph 2-colouring (being almost identical problems). Czumaj and Scheideler give alternative approaches
that allow for non-uniform clause- or edge-sizes \cite{CS00}. We do not investigate these variations more closely as our
current goal is to further improve on the occurrence bound.

The hitherto most powerful approach has been recently published by Srinivasan \cite{Sri08}. It allows to find a satisfying 
assignment to a $k$-CNF in which every variable counts no more than $\mathcal{O}(2^{k/4}/k)$ occurrences. Srinivasan
achieves the improvement by critically augmenting and refining the $2,3$-tree based witness structures used by Alon. 
The resulting algorithm is inherently randomized.

Our contribution bases on definitively getting rid of $2,3$-trees and replacing them by a substantially denser 
witness tree structure. Moreover, we will finally drop the distinction of \emph{bad}, \emph{dangerous}
and \emph{safe} clauses and base the decision of which components to reassign on the witness structures themselves.
As a crucial new aspect of our method, those structures are being made part of the algorithm itself, rather
than just appearing in the probabilistic analysis.

\section{A randomized approach}
\label{randomsection}

\begin{theorem} 
\label{mainrand}
There exists a randomized algorithm that finds a satisfying assignment to 
any $k$-CNF $\varphi$ in which no variable occurs more than $2^{k/2}/(36 k)$ times in expected 
time polynomial in $|\varphi|$.
\end{theorem}

Let $n$ be the number of variables and $m$ the number of clauses of $\varphi$. Let any arbitrary
ordering be imposed on the clauses of $\varphi$ (we will refer to it as the \emph{lexicographic} ordering).
Construct the multigraph $G[\varphi]$ in which every clause of $\varphi$ is a vertex and
any pair of identical or complementary literals in two distinct clauses induces an edge.

We now describe a randomized algorithm that will find a satisfying assignment to $\varphi$. 
Just as in all the previous approaches based on the principle due to Beck and 
Alon, the algorithm will select a (random) preliminary assignment and then solve locally bounded residual
components around dissatisfied clauses. In contrast to the said approaches, our algorithm will
select these components much more restrictively and it will actively construct a collection of
witness trees that allow for a precise probabilistic analysis.

In the description of Algorithm \ref{algloccomsol}, 
a \emph{(primary) witness tree} $T$ is simply a subtree of $G[\varphi]$ with a 
designated root vertex. 

\begin{algorithm}
\caption{Local Component Solver}
\label{algloccomsol}
\begin{algorithmic}
\REQUIRE the formula $\varphi$
\ENSURE a satisfying assignment $\alpha$ to $\varphi$

\algorithmicphase{STEP I: PRELIMINARY ASSIGNMENT}
\vspace*{-0.2cm}

\begin{itemize}
 \item[\textbf{1}] Select a preliminary assigment $\alpha \in \{0,1\}^{\mbox{var}(\varphi)}$ uniformly at random.
\end{itemize}
\vspace*{-0.2cm}

\algorithmicphase{STEP II: CONSTRUCTION OF WITNESS TREES}

Initialize an empty queue $Q$ of clauses and an empty collection of primary witnesses
and then:
\begin{itemize}
 \item[\textbf{2}] select the lexicographically first \textbf{dissatisfied} clause in $\varphi$ as the root of a new primary witness tree $T$.
           Enumerate all the neighbours of the clause in the canonical lexicographic ordering and enqueue each of them in $Q$ 
           unless it is already a member of some tree
 \item[\textbf{3}] dequeue the next clause $D$ from $Q$. If
           \begin{itemize}
             \item[i.]   $D$ contains \textbf{at least $k/2$ variables} that do not yet occur in \textbf{any} tree \textbf{and} 
              \item[ii.] \textbf{all} literals in $D$ over variables not yet occuring in any tree are dissatisfied by $\alpha$,
           \end{itemize}
           then add $D$ to the tree (attaching it in the natural way to the parent by which is was enqueued).        
           Enumerate all the neighbours of the clause in the canonical lexicographic ordering and enqueue each
           of them in $Q$ that is not yet member of any tree.
           If $D$ does not satisfy the requirements, simply skip it (note that it might be enqueued again later).
 \item[\textbf{4}] if $Q$ is non-empty, repeat (3), go to the next step once $Q$ is exhausted
 \item[\textbf{5}] if there is any \textbf{dissatisfied} clause left in $\varphi$ that is neither a member of nor a neighbour to any tree yet, jump back
           to (2), starting construction of a new primary witness using this clause as the root (if there are multiple
           candidates use the lexicographically first one); pass on to (6) once no dissatisfied non-neighbour is available anymore
\end{itemize}

\algorithmicphase{STEP III: DISSECTION}

Let us hereafter call a variable $x$ \emph{covered},
if there exists a witness tree $T$ in the collection
built such that $x$ occurs in a clause included in that tree. We say that a literal is
\emph{covered} if the underlying variable is covered.
\begin{itemize}
 \item[\textbf{6}] inspect every clause of the formula that has not yet been added to any tree. Distinguish:
           if the clause has \textbf{any satisfied literal over a variable not covered} by the witness collection built, 
           delete it (completely, from the formula).
           If it doesn't, truncate it to contain \textbf{only literals over covered variables}. After all clauses
           have been processed, only covered variables are left in the resulting formula $\varphi'$.
\end{itemize}

\algorithmicphase{STEP IV: LOCAL EXHAUSTIVE ENUMERATION}
\begin{itemize}
 \item[\textbf{7}] inspect the connected components of $G[\varphi']$. If any of them contains \textbf{at least $k \log(4m)$ variables},
           cancel the current run and \textbf{restart} from the beginning, sampling another preliminary assignment $\alpha$.
           Otherwise, enumerate all assigments for every component \textbf{exhaustively}, stop at a satisfying one and
           locally replace $\alpha$ by that assignment.
\end{itemize}

\end{algorithmic}
\end{algorithm}

For the performance analysis to follow, let a collection $\mathcal{T}$ of witness trees be given (just
as constructed in the algorithm). Denote $V(\mathcal{T}) := \cup_{T\in\mathcal{T}} V(T)$ and
$\mbox{var}(\mathcal{T}) := \cup_{D\in V(\mathcal{T})} \mbox{var}(D)$.
Let the \emph{natural ordering} $\pi$ of $V(\mathcal{T})$ 
be defined as follows: the vertices $V(T)$ of each tree $T \in \mathcal{T}$ appear consecutively 
in $\pi$ and they are ordered in a level-by-level fashion just as in a BFS starting at the root vertex, 
where siblings are 
ordered lexicographically. The trees among each other appear according to the lexicographic
ordering of their root vertices. Note that the natural ordering is exactly the one in which the
algorithm adds vertices to the tree collection it constructs and yet $\pi$
is fully determined by the shape of $\mathcal{T}$; it is not necessary to consider the construction
history.

For a clause $D \in V(\mathcal{T})$ and a variable $x \in \mbox{var}(D)$, we say that $x$ is 
\emph{novel} in $D$ w.r.t. $\mathcal{T}$ if $D$ is the first clause according to $\pi$ in 
which $x$ occurs. A literal is \emph{novel} in a clause if the underlying variable is novel
in that clause.

Let us now define a \emph{composite witness} $W = \langle \mathcal{T}, V_g\rangle$ 
as a collection $\mathcal{T}$ of vertex-disjoint witness trees together with a
set $V_g \subseteq V(G[\varphi])\backslash V(\mathcal{T})$ of extra vertices such that the following properties are satisfied
\begin{itemize}
 \item[i.] $|\mathcal{T}| > |V_g|$
 \item[ii.] the induced subgraph $G[\varphi][V(\mathcal{T}) \cup V_g]$ is connected
 \item[iii.] every clause $D \in V(\mathcal{T})$ contains at least $k/2$ novel variables. If $D$
             is the root vertex of its tree, then all of its $k$ variables are novel in $D$.
\end{itemize}

We define the \emph{size} $|W|$ of a composite witness $W = \langle \mathcal{T}, V_g\rangle$
to be $|V(\mathcal{T})| + |V_g|$.

We say that a composite witness \emph{occurs} w.r.t. an assignment $\alpha$, if all the novel
literals that appear along $\pi$ are dissatisfied by $\alpha$.

\prbl
\begin{lemma}
\label{algconswit}
Let $\mathcal{T}_0$ be the collection of witness trees that Algorithm \ref{algloccomsol}
constructs, given $\varphi$ and $\alpha$. For each connected component $C$ left in $G[\varphi']$,
there is a composite witness in $G[\varphi]$ that occurs w.r.t. $\alpha$ and which contains 
all the variables in $C$.
\end{lemma}

\proof Let $V = \mbox{var}(C)$ be the set of variables in a given connected component of $G[\varphi']$.
Note that by construction of $\mathcal{T}_0$ and $\varphi'$,
each tree $T \in \mathcal{T}_0$ either lies completely in $C$ and $\mbox{var}(T) \subseteq V$ or it
is completely disjoint, $\mbox{var}(T) \cap V = \emptyset$.\par
Let now $\mathcal{T} \subseteq \mathcal{T}_0$ be the set of witness trees which lie inside $C$. Consider
the subgraph $G[\varphi][V(\mathcal{T})]$ induced by the vertices of $\mathcal{T}$. If this subgraph
is not connected, let us consider all the vertices that are not part of any tree and let us greedily 
add some of them to $V_g$ so as to make $G[\varphi][V(\mathcal{T}) \cup V_g]$ connected. We add
a vertex if and only if it will help to decrease the number of connected components, thus $|V_g|$ is certainly
smaller than the initial number of components and therefore smaller than $|\mathcal{T}|$. Hence, $\langle \mathcal{T},
V_g \rangle$ satisfies properties (i) and (ii) of a composite witness. \par
To check that it satisfies (iii), recall that the natural ordering $\pi$ is the 
same ordering as the one in which Algorithm \ref{algloccomsol} adds clauses to $\mathcal{T}$. The algorithm
will add a clause exlusively if there are at least $k/2$ not-yet-encountered variables contained in it and
all of the not-yet-encountered literals are dissatisfied. This implies not only (iii), but also that
the composite witness built occurs w.r.t. $\alpha$. Note that the fact that the algorithm considers all
trees in $\mathcal{T}_0$ previously collected and not only the ones in $\mathcal{T}$ does not influence
which literals are to be classified novel since all the trees in $\mathcal{T}_0 \backslash \mathcal{T}$ are completely disjoint
in variables from the ones in $\mathcal{T}$. This concludes the argument.\qed

\prbl
\begin{lemma}
\label{witnesscount}
In $G[\varphi]$, there exist at most $m \cdot (2^{k/2}/2)^u$ 
composite witnesses of size exactly $u$.
\end{lemma}

\proof We set out by the observation that there is an injection from
the set of composite witnesses $\langle \mathcal{T}, V_g \rangle$ 
of size $u$ into the set of triples $\langle T, c_v, c_e \rangle$ where
$T$ is an unrooted subtree of $G[\varphi]$ containing $u$ clauses, $c_v$ is a 3-colouring
(or simply a partition into 3 classes) of the vertices $V(T)$ and $c_e$ a 2-colouring of
the edges $E(T)$. We can build such a tree
on top of the vertex set $V(\mathcal{T}) \cup V_g$ by greedily adding edges to the ones
already present in $\mathcal{T}$ until a spanning tree of the connected induced subgraph
$G[\varphi][V(\mathcal{T}) \cup V_g]$ is obtained. Distinguish the edges newly added from the
original ones by colouring them differently. Moreover, use $c_v$ to distinguish the vertices
that originate from $V_g$, the root vertices of the trees in $\mathcal{T}$ and the remainder
of the vertices from one another. It is obvious that the original composite witness can
be reconstructed from the triple. Therefore it is enough to upper bound the number of such triples.\par
According to a simple counting exercise by Donald Knuth \cite{Knu69}, the infinite labelled and rooted
$d$-ary tree has fewer than $(ed)^u$ distinct rooted subtrees of size $u$. Consider now the
$(2^{k/2}/36)$-ary such tree. It has fewer than $(2^{k/2}/12)^u$ many distinct rooted subtrees with $u$ clauses.
Picking
any of them and picking any vertex of $G[\varphi]$ as the root, a subtree $T$ of size $u$ is fully determined:
we start at the selected root and follow the edges that correspond (where the correspondence
is such that the first child corresponds to the (lexicographically) first neighbour in the graph and so forth)
to the ones included in the selected subtree (consider that $G[\varphi]$ has maximum degree smaller than $2^{k/2}/36$).
Additionally, we pick a two-colouring of the edges and a three-colouring of the vertices
for which we have in total fewer than $6^u$ choices. This yields the desired upper bound. \qed

\prbl
\begin{lemma} 
\label{probwitocc}
When sampling $\alpha$ u.a.r., with probability at least $1/2$, no connected 
component with $k \log(4m)$ or more variables is left in $G[\varphi']$. Therefore, the algorithm needs
to jump back no more than 2 times in the expected case.
\end{lemma}

\proof By Lemma \ref{algconswit}, a connected component of $k\log(4m)$ or more variables implies the
existence of a composite witness containing all those variables which occurs w.r.t. $\alpha$.
Such a witness needs to have size at least $\log(4m)$. Let us denote by $X_u$ the random variable
that counts the number of composite witnesses of size exactly $u$ which occur w.r.t. $\alpha$. 
A composite witness $\langle \mathcal{T}, V_g \rangle$ has $|\mathcal{T}|$ root vertices in which
each literal is novel and $u-|V_g|-|\mathcal{T}|$ vertices with at least $k/2$ novel literals each. This
makes a total of $(u-|V_g|-|\mathcal{T}|)\frac{k}{2} + |\mathcal{T}|k = (u-|V_g| + |\mathcal{T}|)\frac{k}{2} > \frac{uk}{2}$
novel literals. For the witness to occur it is required that all of them be dissatisfied.
Hence, a fixed composite witness of size $u$ occurs with probability less than
$2^{-uk/2}$. Applying Lemma \ref{witnesscount}, this yields, by linearity of expectation, $\mathbb{E}[X_u] < m \cdot 2^{-u}$.
Again by linearity, the expected total number $X$ of composite witnesses of size $u$ or larger that occur w.r.t. $\alpha$ is bounded by
$$\mathbb{E}[X] = \sum_{u=\log(4m)}^m \mathbb{E}[X_u] < m \left( \frac{1}{2} \right)^{\log(2m)} \cdot 
\sum_{u=1}^{m-\log(2m)} \left( \frac{1}{2} \right)^u < \frac{1}{2}.$$
Thence, in at least half of the cases, no such witness occurs at all. \qed

Once there are only logarithmically small components, it is obvious that the algorithm can enumerate
all assignments to them in polynomial time. The only thing left to prove is that this will in all
cases yield a satisfying assignment. To this end, the following lemma demonstrates that each clause
left in $\varphi'$ has size at least $k/2$. By \cite{KST93}, a $k/2$-CNF with every variable
occurring at most $2^{k/2}/(36 k)$ times is always satisfiable. Therefore, the algorithm will
find a satisfying assignment by exhaustive enumeration.

\prbl
\begin{lemma} 
\label{clausesize}
Every clause $D \in \varphi'$ has size $|D| \ge k/2$.
\end{lemma}

\proof
Clauses can only be smaller than
$k$ if they have been truncated at some point. Assume that $D$ is a truncated clause.
 Let $D_0 \supseteq D$ be the same clause before truncation.
Did $D_0 \backslash D$ contain any satisfied literal, then it
would have been deleted instead of truncated, therefore all literals of $D_0 \backslash D$
are dissatisfied. If $D$ were empty, then $D_0$ would have been made the root of a new tree, 
therefore $D_0$ is in the neighbourhood of one of the trees constructed by the algorithm and it was 
therefore enqueued into $Q$ at some point (maybe multiple times). Consider
the point in time when $D_0$ was dequeued from $Q$ for the very last time in history.
At that point in time, $D_0$ had exactly the same covered and uncovered variables as it
has now after termination of the algorithm (since any new covering would have 
newly triggered enqueueing). And since all of them are dissatisfied, $D_0$
having more than $k/2$ uncovered literals would be a contradiction since then it would
have been added to the tree at the time of dequeueing. We conclude that $D$ has
to have size at least $k/2$. \qed

\prbl
This concludes the proof of Theorem \ref{mainrand}. \qed

\section{A deterministic alternative}

In this section we will show that there is nothing inherently random to the algorithm exhibited;
it can be easily derandomized. The derandomization technique is in complete analogy to what Beck used
in \cite{Bec91}.

\begin{theorem}
\label{maindet}
There exists a deterministic algorithm that finds a satisfying assignment to 
any $k$-CNF $\varphi$ in which no variable occurs more than $2^{k/2}/(36 k)$ times in 
time polynomial in $|\varphi|$.
\end{theorem}

The following is a well-known fact.

\prbl
\begin{lemma}
\label{algsmallexp}
Let $\psi$ be any CNF and let $X(\psi)$ denote the random variable
counting the number of dissatisfied clauses under an assignment sampled u.a.r. If $\mathbb{E}[X(\psi)] < 1$,
then $\psi$ is satisfiable and a satisfying assignment can be found in time polynomial in $\psi$.
\end{lemma}

\proof Simply pick any variable $x \in \mbox{var}(\psi)$ and 
have the algorithm calculate both $\mathbb{E}[X(\psi)|^{x=0}]$ and $\mathbb{E}[X(\psi)|^{x=1}]$.
Since the two values must average to $\mathbb{E}[X(\psi)] < 1$, at least one of them is smaller than $1$.
We assign the corresponding value to $x$ and continue recursively. \qed

We shall now demonstrate that it is sufficient to consider witnesses in a certain size
range to exclude all large witnesses.

Let $W = \langle \mathcal{T}, V_g \rangle$ and $W' = \langle \mathcal{T}', V_g' \rangle$ be composite witnesses
in $\varphi$. We say that $W$ \emph{implies} $W'$ if for all $\alpha$ for which $W$ occurs, $W'$ occurs as well.

\begin{lemma}
\label{impliedwitnesses}
Let $u>k>2$. For every composite witness $W$ in $\varphi$ with $|W| \ge u$, there
exists a witness $W'$ in $\varphi$ having $u \le |W'| \le (k+1) u$ such that $W$ implies $W'$.
\end{lemma}

\proof Assume the contrary and let, for fixed $u$ and $k$, $W=\langle \mathcal{T}, V_g \rangle$ 
be a smallest counterexample to the claim, so let $W$ be a composite witness of size at least $u$ 
which does not imply any witness $W'$ with a size in the range $u \le |W'| \le (k+1) u$. We observe that $W$
must be of size larger than $(k+1) u$ since otherwise $W'=W$ would fulfill the requirements.\par
Suppose $V_g \ne \emptyset$. Pick any arbitrary clause $D \in V_g$ and remove it from $V_g$,
building $W' := \langle \mathcal{T}, V_g \backslash \{D\} \rangle$. If $W'$ still is a composite witness,
we have found a smaller counterexample which is a contradiction. Therefore it must hold that 
$H := G[\varphi][V(\mathcal{T}) \cup V_g \backslash \{D\}]$ is no longer connected. By the structure of $G[\varphi]$
and since $D$ contains exactly $k$ variables, $H$ cannot have more than $k$ connected components of which the
largest component $C_L$ must have size at least $(|W|-1)/k \ge u$. If we now select all trees and all vertices
from $V_g$ that lie inside $C_L$, then this constitutes a smaller composite witness of size at least $u$, which
either has the desired properties or is a smaller counterexample (it can easily be checked that all properties
that define a composite witness are preserved; most notably, the number of vertices in $V_g \cap V(C_L)$ is smaller
than the number of trees inside $C_L$ since otherwise we would have a superfluous node, removal of which would not disconnect
the structure and hence it would have been spare in $W$ already, contradicting minimality).\par
Suppose $V_g=\emptyset$. Let us pick the very last clause in $\mathcal{T}$ according to the natural
ordering and let us remove that clause from its tree (if the tree had size 1, simply delete it). By the very same
argument as before, the remainder either stays connected or falls apart into at most $k$ connected components, the
largest of which contains a composite witness of size at least $u$. And since we have removed the very last clause,
all novel literals in the preserved clauses remain novel and so that witness occurs as well. \qed

\prbl
\begin{lemma}
\label{bigformula}
Associate with every composite witness $W$ a clause $D_\varphi(W)$ which consists of all novel
literals that occur alongside the natural ordering. Let $\Psi_j$ denote the set of all composite witnesses  
of size exactly $j$ in $\varphi$. Build the formula 
$$\psi := \bigcup_{j = \log(2m)}^{(k+1)\log(2m)} D_\varphi[\Psi_j].$$
Then $\psi$ is satisfiable, a satisfying assignment $\alpha$ to it can be found in time polynomial in $|\varphi|$ and
w.r.t. such an $\alpha$, no composite witness of size $\log(2m)$ or larger occurs in $\varphi$.
\end{lemma}

\proof Let $X_j$ denote the random variable counting the number of dissatisfied clauses in the formula $D_\varphi[\Psi_j]$ when sampling
truth assignments uniformly at random. According to our earlier considerations, we have $|\Psi_j| < m \cdot (2^{k/2}/2)^j$. Every witness in
$\Psi_j$ contains at least $jk/2$ novel literals. Therefore, $\mathbb{E}[X_j] < 2^{-jk/2} \cdot m \cdot (2^{k/2}/2)^j = m 2^{-j}.$ Let now $X$
denote the number of dissatisfied clauses in the formula $\psi$ when sampling assignments to it uniformly at random. We obtain
$$\mathbb{E}[X] = \sum_{j = \log(2m)}^{(k+1)\log(2m)} \mathbb{E}[X_j] < m \cdot 2^{-\log(m)} \cdot \sum_{j = 1}^{k\log(2m)+1} 2^{-j} < 1.$$
By Lemma \ref{algsmallexp}, $\psi$ is satisfiable and a satisfying assignment to it can be found in time polynomial in $|\psi|$, which,
in turn, is polynomial in $|\varphi|$.\par
Now let $\alpha$ be an assignment that satisfies $\psi$. Suppose there is a composite witness $W$ of size
$\log(2m)$ or larger in $\varphi$ that occurs w.r.t. $\alpha$. By Lemma \ref{impliedwitnesses}, there exists $W'$ which occurs as well and for
which we have $\log(2m) \le |W'| \le (k+1) \log(2m)$. This witness is therefore contained in some $\Psi_j$ that we enumerated and
since $\alpha$ satisfies $\psi$, it satisfies $D_\varphi(W')$, so there must be at least one novel literal in $W'$ which
is satisfied. This contradicts occurrence of $W'$. \qed

\noindent
\textit{Proof of Theorem \ref{maindet}.} As in Lemma \ref{bigformula}, 
we first enumerate all composite witnesses $W$ with $\log(m) \le |W| \le (k+1) \log(m)$
and build the respective clauses $D_\varphi(W)$ from them, producing $\psi$. We solve $\psi$ using the designated strategy.
Now that we are sure that no witness of size $\log(2m)$ or larger exists, we can safely run the (deterministic) 
remainder of the algorithm described in Section \ref{randomsection} and we will be sure that all residual connected components 
are of logarithmic size. This will solve $\varphi$ in time polynomial in $|\varphi|$. \qed

\section{Conclusion}

We have demonstrated that $\mathcal{O}(2^{k/2}/k)$-occurrence instances of $k$-SAT are easy to solve.

Still, there remains a big gap
between the class of problems for which the Local Lemma predicts a solution
and the one for which a polynomial-time search algorithm is known to exist. The main open
question is whether this gap can be completely closed or whether there is something inherently
more difficult to the search question as compared to the existence question.

While various aspects of the present solution suggest that some natural threshold has been
hit at $k/2$ and another slight variant of the same method will not bring about another considerable
improvement, the history of the problem demonstrates that conjectures of this sort ought to be
treated with due scepticism.

\section*{Acknowledgements}

Many thanks go to Dominik Scheder for various very helpful comments and to
my supervisor Prof. Dr. Emo Welzl for the continuous support.


\begin{thebibliography}{AEKS81}

\bibitem[Knu69]{Knu69}
Donald E. Knuth. 
\emph{The Art of Computer Programming}, Vol. I, Addison Wesley,
London, 1969, p. 396 (Exercise 11).

\bibitem[EL75]{EL75}
Paul Erd\"os and L\'aszl\'o Lov\'asz.
\emph{Problems and results on 3-chromatic hypergraphs and some related questions}. 
In A. Hajnal, R. Rado and V.T. S\'os, editors,
Infinite and Finite Sets (to Paul Erdos on his 60th brithday), volume II,
pages 609-627. North-Holland, 1975.

\bibitem[Bec91]{Bec91}
J\'oszef Beck.
\emph{An Algorithmic Approach to the Lov\'asz Local Lemma}.
Random Structures and Algorithms, 2(4):343-365, 1991.

\bibitem[Alo91]{Alo91}
Noga Alon. 
\emph{A parallel algorithmic version of the local lemma}.
Random Structures and Algorithms, 2(4):367-378, 1991.

\bibitem[KST93]{KST93}
Jan Kratochv\'il and Petr Savick\'y and Zsolt Tuza.
\emph{One more occurrence of variables makes satisfiability jump from trivial to NP-complete}. 
SIAM J. Comput., Vol. 22, No. 1, pp. 203-210, 1993.

\bibitem[MR98]{MR98}
Michael Molloy and Bruce Reed.
\emph{Further Algorithmic Aspects of the Local Lemma.} 
In Proceedings of the 30th Annual ACM Symposium on the Theory of Computing, pages 524-529, 1998.

\bibitem[CS00]{CS00}
Artur Czumaj and Christian Scheideler. 
\emph{Coloring non-uniform hypergraphs: a new algorithmic approach to the general Lovasz local lemma.} 
Symposium on Discrete Algorithms, 30-39, 2000.

\bibitem[Mos06]{Mos06}
Robin A. Moser.
\emph{On the Search for Solutions to Bounded Occurrence Instances of SAT}.
Not published. Semester Thesis, ETH Z\"urich. 2006.

\bibitem[Sri08]{Sri08}
Aravind Srinivasan.
\emph{Improved algorithmic versions of the Lov\'asz Local Lemma}.
Proceedings of the nineteenth annual ACM-SIAM symposium on Discrete algorithms (SODA),
San Francisco, California, pp. 611-620, 2008.

\end{thebibliography}
\end{document}